\documentclass[onecolumn]{emulateapj}

\usepackage{epsfig}
\usepackage{multirow}
\usepackage{graphics}
\usepackage{amsmath, amsthm, amssymb}

\newcommand     \Msol   {M_{\odot}}

\newcommand{\beq}{\begin{equation}}
\newcommand{\eeq}{\end{equation}}
\newcommand{\beqa}{\begin{eqnarray}}
\newcommand{\eeqa}{\end{eqnarray}}

\newcommand{\gcc}         {\rm g\:cm^{-2}}

\countdef\decade=200
\advance\decade by \year
\countdef\hours=201
\advance\hours by \time
\divide\hours by 60
\countdef\mins=202
\advance\mins by \hours
\multiply\mins by 60
\multiply\hours by 100
\countdef\miltime=203
\advance\miltime by \hours
\advance\miltime by \time
\advance\miltime by -\mins

\begin{document}

\title{The Darkest Shadows:\\Deep Mid-Infrared Extinction Mapping of a Massive Protocluster}


\author{Michael J. Butler}
\affil{Institute of Theoretical Physics, University of Z\"urich, CH-8057 Z\"urich, Switzerland}
\author{Jonathan C. Tan}
\affil{Depts. of Astronomy \& Physics, University of Florida, Gainesville, FL 32611, USA}
\author{Jouni Kainulainen}
\affil{Max-Planck-Institute for Astronomy, K\"onigstuhl 17, 69117, Heidelberg, Germany}

\begin{abstract} 
We use deep $8\:\micron$ Spitzer-IRAC imaging of a massive Infrared
Dark Cloud (IRDC) G028.37+00.07 to construct a Mid-Infrared (MIR) extinction map
that probes mass surface densities up to $\Sigma\sim1\:\gcc$
($A_V\sim200\:$mag), amongst the highest values yet probed by
extinction mapping. Merging with a NIR extinction map of the region,
creates a high dynamic range map that reveals structures down to
$A_V\sim1\:$mag. We utilize the map to: (1) Measure a cloud mass
$\sim7\times10^4\:M_\odot$ within a radius of $\sim8\:$pc.  $^{13}$CO
kinematics indicate that the cloud is gravitationally bound. It thus
has the potential to form one of the most massive young star clusters
known in the Galaxy. (2) Characterize the structures of 16 massive
cores within the IRDC, finding they can be fit by singular
polytropic spheres with $\rho\propto{r}^{-k_\rho}$ and
$k_\rho=1.3\pm0.3$. They have $\overline{\Sigma}\simeq0.1-0.4\:\gcc$ ---
relatively low values that, along with their measured cold temperatures, 
suggest magnetic fields, rather than accretion-powered radiative
heating, are important for controlling fragmentation of these
cores. (3) Determine the $\Sigma$ (equivalently column density or
$A_V$) probability distribution function (PDF) for a region that is
near complete for $A_V>3\:$mag. The PDF is well fit by a single
log-normal with
mean $\overline{A}_V\simeq9\:$mag,
high compared to other known clouds. 
It does not exhibit a separate high-end power law tail, which has been
claimed to indicate the importance of self-gravity. However, we
suggest that the PDF does result from a self-similar, self-gravitating
hierarchy of structure being present over a wide range of scales in
the cloud.
\end{abstract}


\section{Introduction}\label{S:intro}

Most stars, especially massive ones, form in clusters from dense
clumps of gas inside giant molecular clouds \citep[e.g.,][]{mckee2007}.
Turbulence and magnetic fields are thought to
help regulate star formation activity, but both their absolute and
relative importance are uncertain. These can affect the overall
timescale of star cluster formation, be it dynamically fast 
\citep{elmegreen2007}
or slow 
\citep{tan2006},
the fragmentation of the gas into
self-gravitating cores 
\citep{padoan2002,vazquez2005,kunz2009},
and thus the mechanism by which massive stars are born, i.e., via
competitive clump-fed accretion \citep{bonnell2001,wang2010}
or via core accretion 
\citep{mckee2003},
and the stellar initial mass
function established.

Progress requires improved observational constraints on the properties
of dense gas clumps that are on the verge of massive star and star
cluster formation, such as the cold, high column density clouds that
reveal themselves as IRDCs, silhouetted against
the MIR emission from the Galactic diffuse interstellar
medium 
\citep[e.g.,][]{carey1998,rathborne2006,butler2009}.
IRDCs suffer from CO 
freeze out onto dust grains 
\citep[e.g.,][]{hernandez2011}
and thus have typically been studied via their mm to far-IR
(FIR) dust continuum emission 
\citep[e.g.,][]{rathborne2006,peretto2010}.
This method has the disadvantage of requiring knowledge of both the
emissivity and temperature of dust in the cloud.
Single dish observations are needed to recover the total flux, but
these have relatively poor angular resolution (e.g. 11\arcsec\ FWHM
angular resolution for the $1.2\:$mm observations of \citet{rathborne2006}
with the IRAM $30\:$m telescope; $\sim$22\arcsec\ for the
$250\:\micron$ observations of 
\citet{peretto2010}
with {\it Herschel}-SPIRE). Interferometric observations are possible
at submm and longer wavelengths, but on their own provide poor
constraints on dust temperature.


Extinction mapping is a temperature-independent method to probe cloud
structure. However, using background stars in the NIR, clouds are
typically only probed up to $A_V\simeq 25\:$mag\footnote{We interchange
  between $\Sigma$, $N_{\rm{H}}$ and $A_V$ via
  $A_V/(1\:{\rm{mag}})\equiv\:N_{\rm{H}}/(1.9\times10^{21}\:{\rm{cm}^{-2}})\equiv\Sigma/(4.45\times10^{-3}\:\gcc)$ \citep[see][]{kainulainen2013}.} 
\citep{kainulainen2011},
while IRDCs can have column densities
$\sim10\times$ larger. Furthermore, the location of IRDCs at
$\gtrsim\:$kpc distances in the crowded Galactic plane, with NIR
sources at a range of distances, necessitates statistical methods that
limit the effective angular resolution of the maps to $\sim 30\arcsec$.

Thus, MIR extinction (MIREX) mapping has proven more effective at
probing IRDCs, e.g. with 2\arcsec\ angular resolution achieved with
{\it{Spitzer}}-IRAC GLIMPSE \citep{churchwell2009} $8\:\micron$
images \citep[e.g.,][]{butler2009}.
Using the foreground estimation method of \citet{butler2012},
the maximum $\Sigma$ that can be probed depends on the noise level of
the images: a $1\:\sigma$ level of $0.6\:$MJy/sr for GLIMPSE images
leads to a maximum, ``saturation'', mass surface density of
$\Sigma_{\rm{sat}}\sim0.3-0.5\:\gcc$ for typical inner Galaxy
IRDCs. Since it compares specific intensities towards the IRDC with
those of its surroundings, MIREX mapping has difficulties at lower
$\Sigma$ values, where it tends to underestimate true extinctions by
$A_V\sim5-10\:$mag. This deficiency can be fixed by combining MIR and
NIR-derived maps \citep{kainulainen2013}.

Here we present an $8\:\micron$ extinction map of IRDC G028.37+00.07,
hereafter IRDC~C \citep{butler2009}, with kinematic distance of
$5.0\:$kpc \citep{simon2006} made using longer-exposure archival
{\it{Spitzer}}-IRAC data, probing to
$\Sigma_{\rm{sat}}\simeq1\:\gcc$. Combining with a NIR extinction map,
reveals a very high dynamic range of mass surface densities.
 
\section{Methods}\label{S:method}




\citet{butler2012} MIREX mapping requires knowing the intensity of radiation
just behind the cloud, $I_{\nu,0}$,
(estimated by interpolation of surrounding observed intensities)
and just in front, $I_{\nu,1}$. Then for negligible
emission in the cloud and a 1D geometry, $I_{\nu,1}=
e^{-\tau_\nu}\:I_{\nu,0}$, where optical depth
$\tau_\nu=\kappa_\nu\Sigma$, where $\kappa_\nu$ is total opacity at
frequency $\nu$ per unit total mass. However, foreground
emission from the diffuse ISM emission causes us to observe
$I_{\nu,1,{\rm{obs}}}=I_{\nu,{\rm{fore}}}+I_{\nu,1}=I_{\nu,{\rm{fore}}}+e^{-\tau_\nu}\:I_{\nu,0}$ towards the IRDC
and 
$I_{\nu,0,{\rm{obs}}}=I_{\nu,{\rm{fore}}}+I_{\nu,0}$
towards the surroundings used to estimate $I_{\nu,0}$.

\begin{figure*}[!tb]
\begin{center}$
\begin{array}{c}
\includegraphics[width=2.9in]{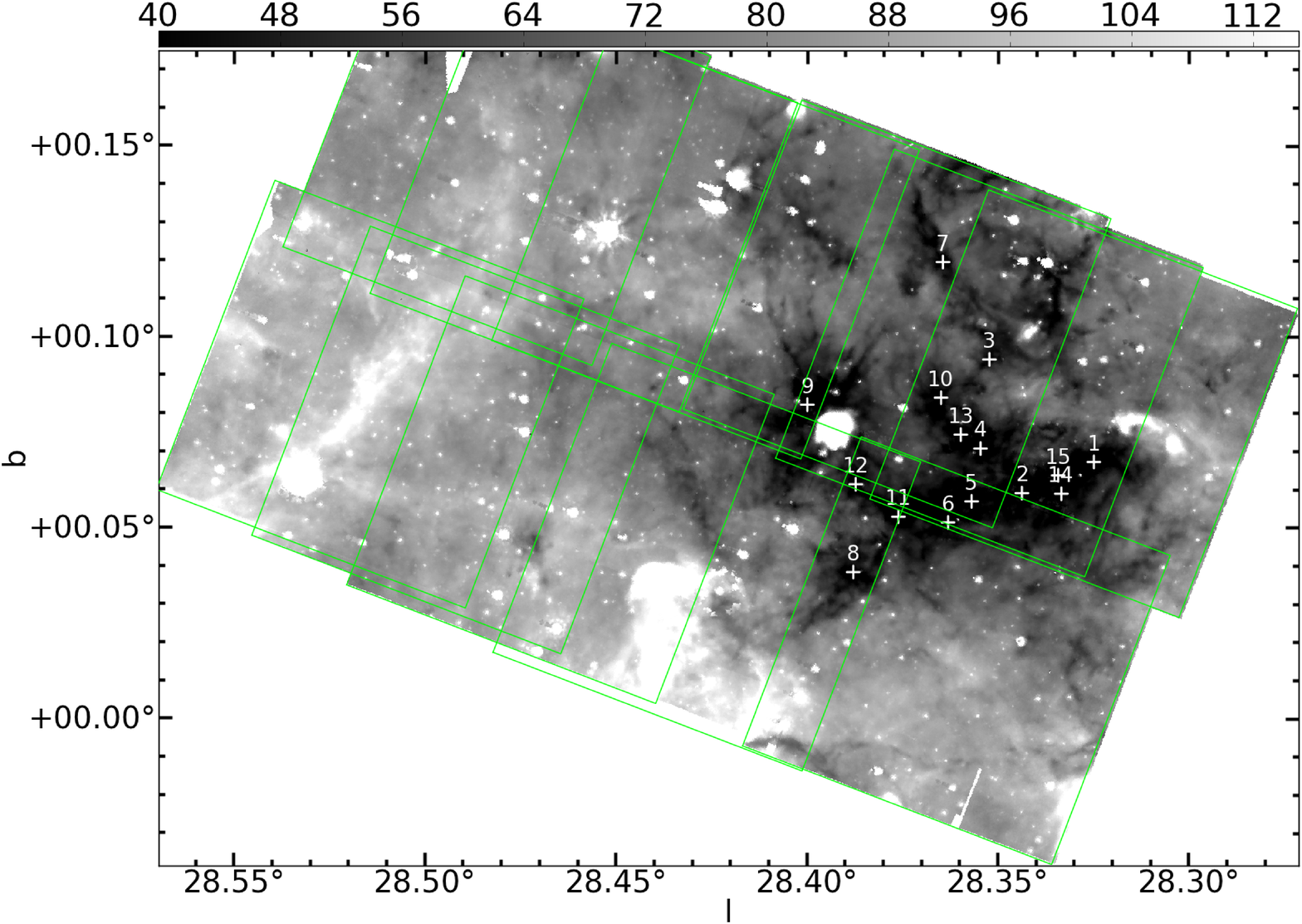}  
\includegraphics[width=2.9in]{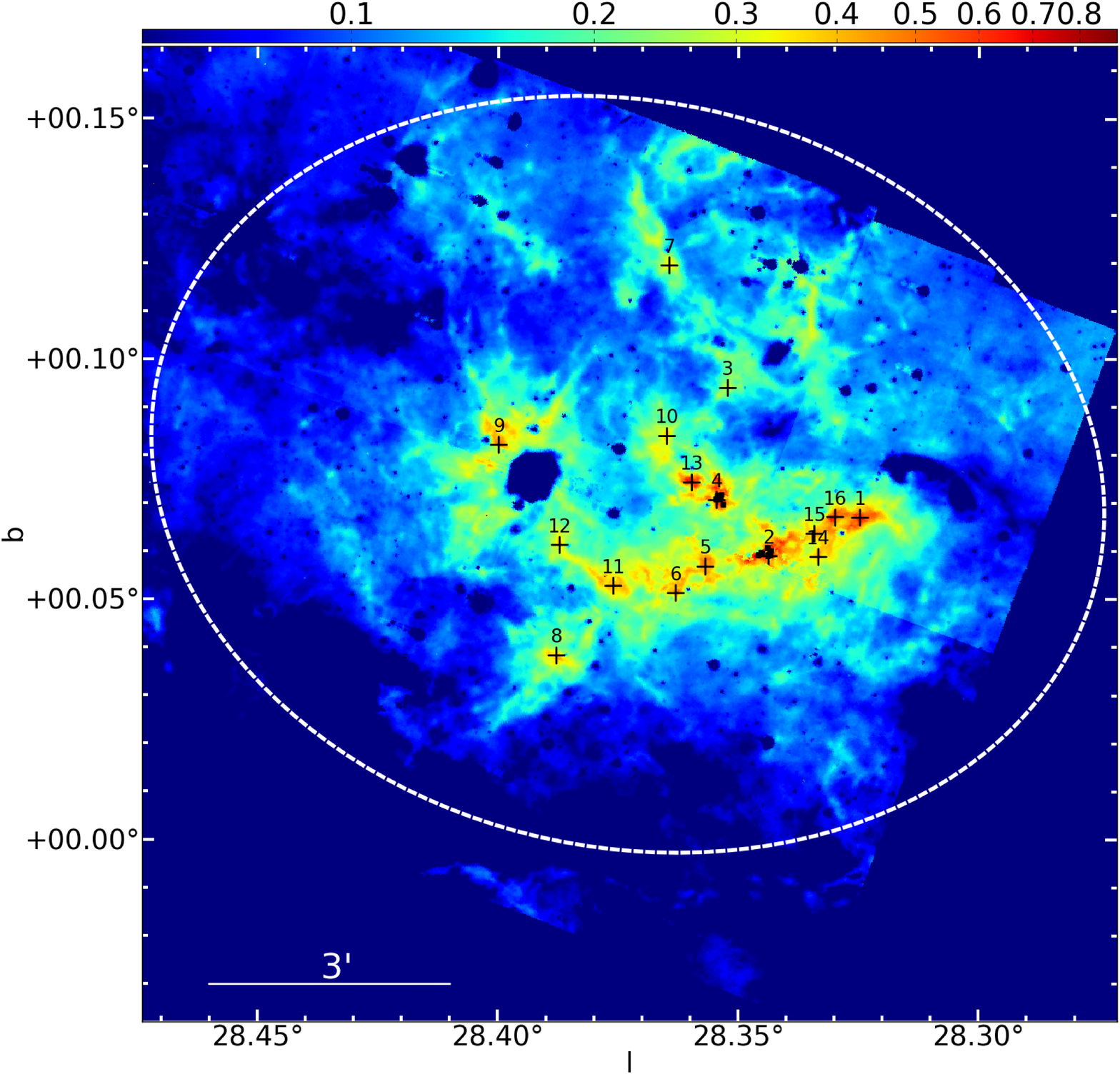}
\end{array}$
\end{center}
\caption{\footnotesize
{\it{(a)~Left}}: $8\micron$ image of IRDC G028.37+00.07 (Cloud C) with
intensity scale in MJy/sr. Dotted squares show the 12 {\it{Spitzer}}
IRAC pointings, each with $10.4\:$s exposure, used to construct the
mosaiced image. {\it{(b)~Right}}: Merged NIR+MIR $\Sigma$ map (units
of $\gcc$). The IRDC-defining ellipse from \citet{simon2006} is shown. Core/clump
centers (\S\ref{S:cores}) are labeled with crosses. Saturated pixels
are marked with white squares.  }
\label{fig:cloudC}
\end{figure*}

Following \citet{butler2012}, we estimate $I_{\nu,{\rm{fore}}}$ by searching for
saturation in independent cores, occurring when $\Sigma$ is large
enough to block essentially all background emission.
\citet{butler2012} used {\it Spitzer} GLIMPSE data with $2.4\:$s exposures.  We now
use archival data of IRDC C (PI G. Fazio; Project ID DARKCLOUDS/219;
AOR key 6050304), with $10.4\:$s exposure per pointing. After starting
with PostBCD data (pipeline version S18.18.0), the final combined
image consists of 12 mosaiced $5.2'\times5.2'$ pointings, resulting in
$\sim20$ regions that have different total exposures, $1\sigma$ noise
and instrumental background levels (Figure~\ref{fig:cloudC}a).
For each region with area $\gtrsim$500 pixels, we
search for saturation by:
(1) Find $I_{\nu,1,{\rm{obs}}}({\rm{min}})$, the minimum value of
$I_{\nu,1,{\rm{obs}}}$.  (2) Find all pixels with
$I_{\nu,1,{\rm{obs}}}({\rm{min}})<I_{\nu,1,{\rm{obs}}}<I_{\nu,1,{\rm{obs}}}({\rm{min}})+2\sigma$. 
If these are spatially independent 
(i.e., extended over $\geq4\arcsec$, cf. $8\arcsec$ of \citet{butler2012}), then the
region is defined to exhibit ``local saturation'' and all pixels in
this intensity range are labeled ``saturated''.  Steps 1 and 2 are
repeated for each region. (3) Evaluate the mean value of
$I_{\nu,1,{\rm{obs}}}$ of saturated pixels in all regions,
$I_{\nu,1,{\rm{obs}}}({\rm{sat}})$ (in practice only 2 regions exhibit
saturation, see \S\ref{S:cores}). Apply an offset to each saturated
region equal to the difference between
$I_{\nu,1,{\rm{obs}}}({\rm{sat}})$ and its local value. We expect
these offsets result from varying instrumental background noise.
We then subtract a local $2\sigma$ intensity from the foreground to
each region, which ensures every pixel has a finite estimate of
$\Sigma$. The average estimate of $I_{\rm \nu,fore}$ is
$\simeq31.2\:$MJy/sr ($\simeq0.6\:$MJy/sr greater than the \citet{butler2012} value).
Lack of areal coverage of the archival data means we must use the
background model derived from GLIMPSE \citep{butler2012}. This was scaled to match
intensities of the archival data by comparing median intensities in
several small patches free of stellar or extended emission sources,
which were used to derive a single mean offset factor.

Longer integration times (everywhere $\geq10.4\:$s; all identified
core/clumps [\S\ref{S:cores}] have $\geq20.8\:$s, up to maximum of
$52\:$s; compared to $2.4\:$s for GLIMPSE), cause the $1\sigma$ noise
level to fall from $\sim0.6\:$MJy/sr \citep{reach2006}
to $0.29,0.20,0.14\:$MJy/sr for the $10.4,20.8,41.6\:$s exposure
regions, respectively. While the absolute value of this noise level is
somewhat uncertain, the relative values should be better determined.
By subtracting a $2\sigma$ noise level, our method sets the minimum value of
$I_{\nu,1}\simeq1-2\sigma$, leading to a limit at which our $\Sigma$ measurements
begin to be underestimated by saturation effects,
$\Sigma_{\rm{sat}}=\tau_{\nu,{\rm{sat}}}/\kappa_{\nu}={\rm{ln}}(I_{\nu,0}/I_{\nu,1})/\kappa_{\nu}$,
of 
$0.625,0.675,0.722\:\gcc$ 
respectively, assuming $I_{\nu,1}=2\sigma$ and a value of
$I_{\nu,0}=I_{\nu,0,{\rm{obs}}}-I_{\nu,{\rm{fore}}}=(94.3-31.2=63.1)\:$MJy/sr,
the mean intensity of our scaled background model with foreground
subtracted. This represents an increase of up to $\sim$40\% in dynamic
range compared to the \citet{butler2012} map, which had
$\Sigma_{\rm{sat}}\simeq0.5\:\gcc$.
A true $\Sigma$ value that is equal to $\Sigma_{\rm{sat}}$
is underestimated by about 7\% \citep[cf. 17\% for][]{butler2012}.
Thus the new map not only probes to higher $\Sigma$, but does so with
greater accuracy. Note, values of $\Sigma>\Sigma_{\rm{sat}}$ are
present in the map, up to $\sim0.85\:\gcc$.

The above equation for $\Sigma_{\rm{sat}}$ is also useful for seeing
the effect of errors in estimating the background specific intensity,
$I_{\nu,0}$. This is expected to show 
fluctuations about the median interpolated level: \citet{butler2009} find the HWHM
of the distribution of background intensities around IRDC C to be a
factor of 1.2. This introduces an error of
$\Delta\Sigma=({\rm{ln}}\:1.2)/\kappa_\nu\rightarrow0.024\:\gcc$,
which is only a $\sim3-4\%$ error for $\Sigma\simeq\Sigma_{\rm{sat}}$,
growing to 20\% in the optically thin limit.
 
In several locations, $\Sigma$ ``holes'' are seen, where $\Sigma$ is
significantly lower than the surrounding pixels. This could be caused
by the presence of a real MIR-bright point source or by an
instrumental artifact, such as a bad pixel.  To identify the latter,
we use the original GLIMPSE $8\micron$ image as a reference.  If there
is no source in the GLIMPSE image, then the hole in the new image is
corrected by replacing the bad pixel with the mean of the nearest
unaffected pixels.  This typically occurs far from any of the centers
of our core sample (\S\ref{S:cores}), with the exception of C1, which
had a hole directly adjacent to core center.  


Finally, we follow \citet{kainulainen2013} to merge the MIREX map with a NIR extinction
map of the region, derived using the method of \citet{kainulainen2011} and data from the
\emph{UKIRT/Galactic Plane Survey} \citep{lawrence2007}.
The technique estimates NIR extinction towards background stars by
comparing their NIR colors to those of stars in a nearby reference
field.
The zero-point calibration uncertainty is estimated to be $A_V\sim1\:$mag.  
To combine the NIR and MIR extinction maps, a relative dust
opacity-law of $\tau_{8\micron}=0.29\tau_K$ is adopted \citep{kainulainen2013}.
The net effect of the combination is a zero point shift of the MIREX
map by $A_V\sim5-10\:$mag, which varies at the resolution of the NIR
map of 30\arcsec. In regions where $A_V\gtrsim15\:$mag, the NIR map is
not accurate so we interpolate its value from surrounding regions. The
final combined map is shown in Figure~\ref{fig:cloudC}b.

\section{Results}

\subsection{Overall Cloud Mass and Dynamics}\label{S:dynamics}

Inside the IRDC elliptical boundary of \citet{simon2006} (effective
radius of $R_{\rm{eff}}=7.7\:$pc, $e=0.632$) the mean mass surface density is
$\overline{\Sigma}=0.0926\gcc$ (ignoring regions affected by MIR-bright
sources). We assume a 30\% uncertainty due to the opacity per unit
mass, which includes dust opacity and dust-to-gas mass ratio
uncertainties \citep{butler2012}. This result is 28\% higher than the value of
$\overline{\Sigma}=0.0721\:{\rm g\:cm^{-2}}$ from \citep{kainulainen2013}, which we attribute
to the new map's ability to probe to higher values of $\Sigma$.

Given $\overline{\Sigma}$ and the kinematic distance of $5.0\:$kpc, for
which we assume 20\% uncertainty, the total IRDC mass inside the
ellipse is $M=6.83\times10^4\:M_\odot$, with 50\% overall
uncertainty. This compares with our previous estimate of
$5.32\times10^4\:M_\odot$ \citep{kainulainen2013}.
This shows that IRDC C is one of the more massive molecular clumps in
the Galaxy. Its mass is comparable to the $1.1\:$mm Bolocam Galactic
Plane Survey (BGPS) clumps of \citet{ginsburg2012} and the
$1.3\times10^5\:M_\odot$, $2.8\:$pc-radius Galactic center clump (the
``Brick'') of \citet{longmore2012}. Note that these masses are based
on dust continuum emission and thus suffer from additional
uncertainties due to dust temperature estimates. With this caveat in
mind, these massive mm clumps appear to be a factor of about 10 denser
than IRDC C, which, inside $R_{\rm eff}$, we estimate to be
$n_{\rm{H}}\simeq1120\:{\rm{cm}^{-3}}$ (but with the caveat of
assuming spherical geometry), corresponding to a free-fall time of
$t_{\rm{ff}}=[3\pi/(32G\rho)]^{1/2}=1.30\times10^{6}\:$yr.  Apart from
the Brick, the mm clumps are typically already undergoing active star
formation. Some stars have started to form in IRDC C \citep[e.g. the
  MIR-bright sources; see also][]{zhang2009}, but overall the star
formation activity appears to be relatively low, and so this cloud
should represent an earlier stage of massive star cluster formation.

%
%
From $^{13}{\rm{CO}}(1-0)$ Galactic Ring Survey data
\citep{jackson2006}, the IRDC's 1D velocity dispersion, $\sigma$,
(integrating over the \citet{simon2006} ellipse) is $3.41-3.75\:{\rm{km\:s^{-1}}}$
\citep[depending on method of gaussian fitting or total integration,][]{kainulainen2013}.
If the cloud is virialized, then $\sigma$ should be
$4.2\:{\rm{km\:s^{-1}}}$ (ignoring surface pressure and magnetic field
terms and evaluated for a spherical cloud with internal power law
density profile of $\rho\propto{r}^{-k_{\rho,{\rm{cl}}}}$ with $k_{\rho,{\rm{cl}}}=1$;
Bertoldi \& McKee 1992). Support by large scale magnetic fields would
reduce the virial equilibrium velocity dispersion, although such
support is unlikely to be dominant given an expected field strength of
$12\:{\rm{\mu}\rm{G}}$ implied by the cloud's mean density (Crutcher et
al. 2010). Thus, overall the cloud appears close to virial equilibrium
(or perhaps sub-virial), implying it is gravitationally bound and that
self-gravity is important over the largest size scales defining the
cloud.

\subsection{Internal Structure of Cores and Clumps}\label{S:cores}

\begin{figure*}[!tb]
\begin{center}$
\begin{array}{ccc}
\hspace{-0.0in} 
\includegraphics[width=1.9in]{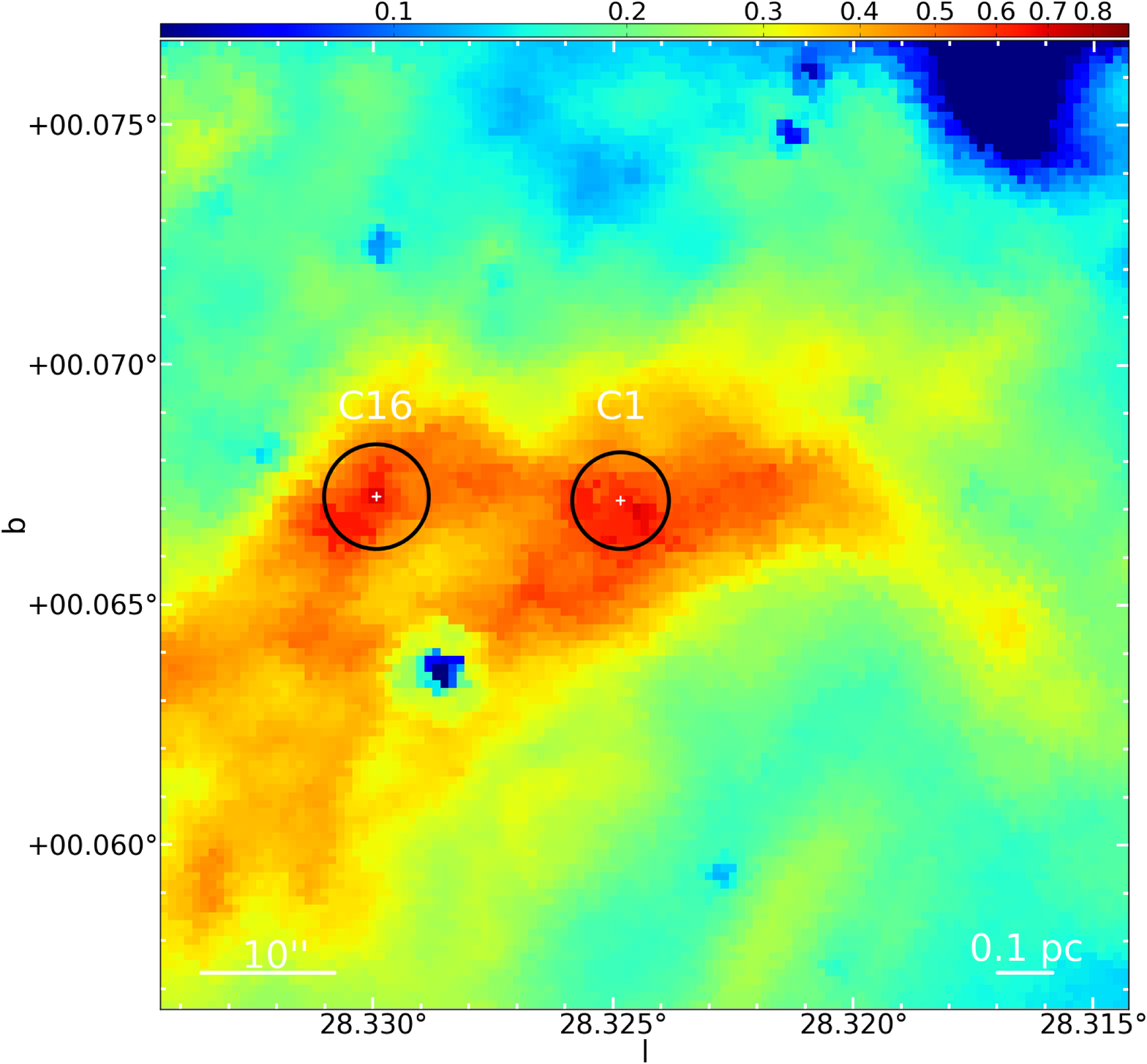} & 
\hspace{-0.0in} 
\includegraphics[width=1.9in]{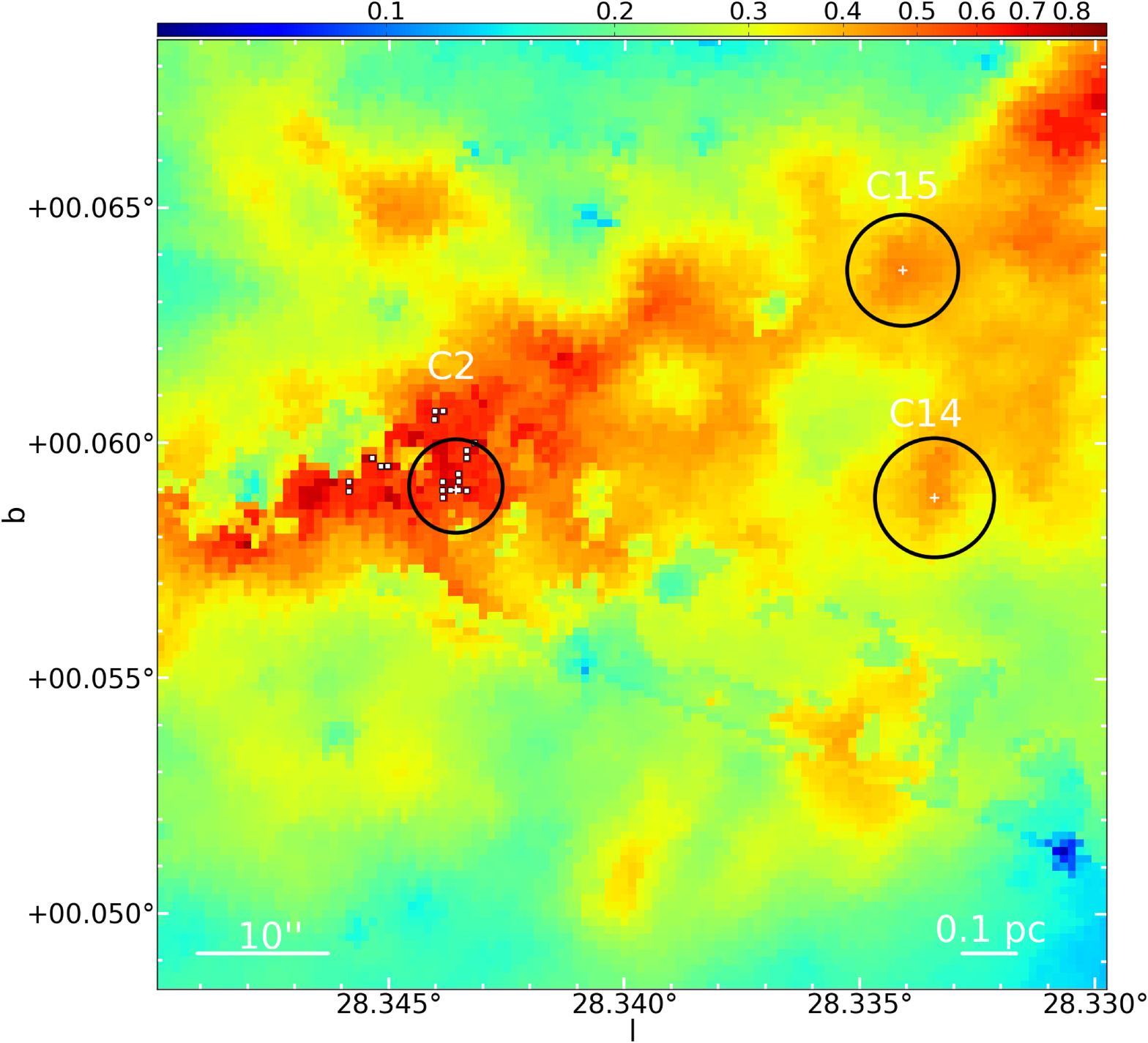} & 
\hspace{-0.0in} 
\includegraphics[width=1.9in]{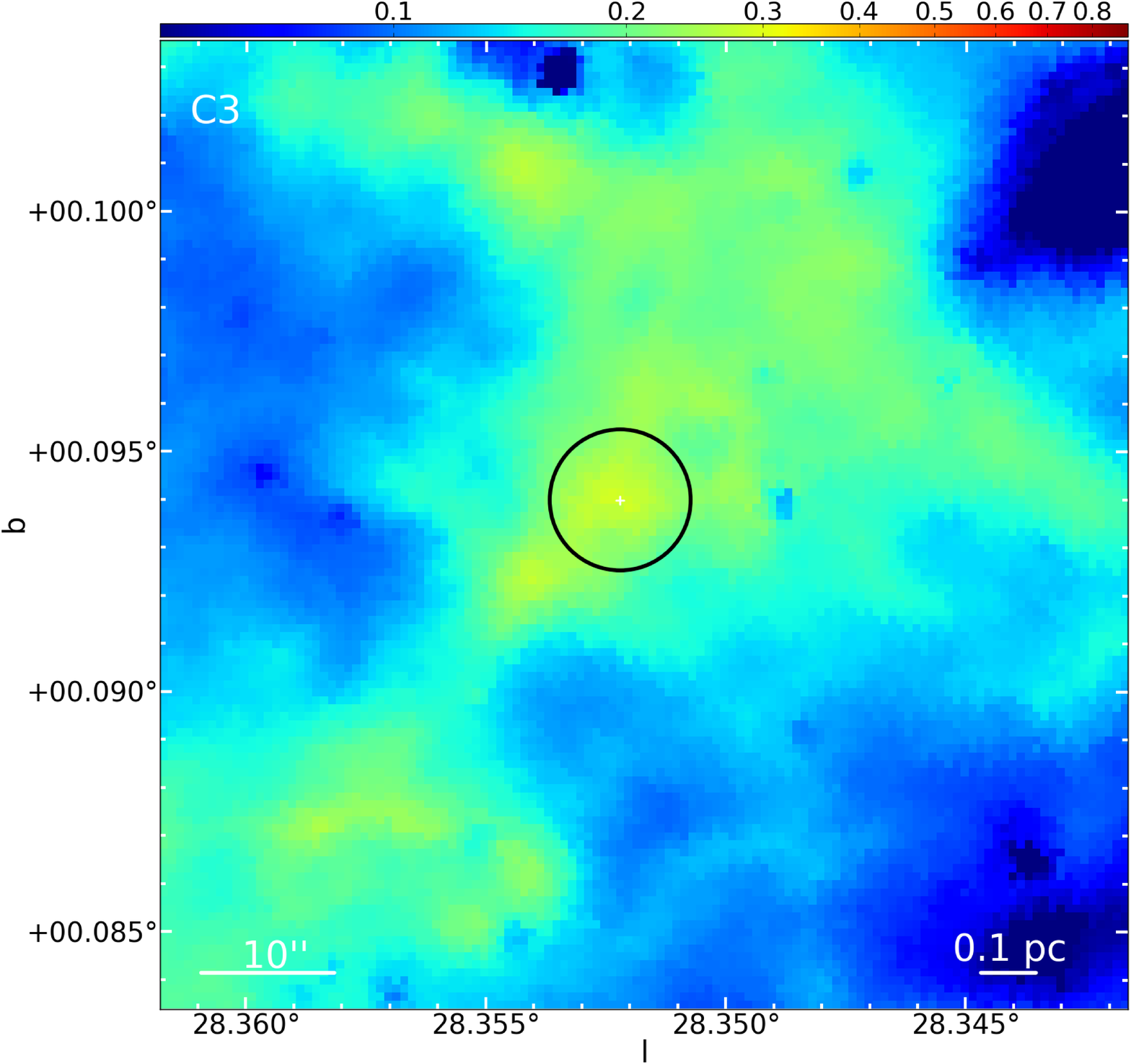} \\
\hspace{-0.0in} \includegraphics[width=1.9in]{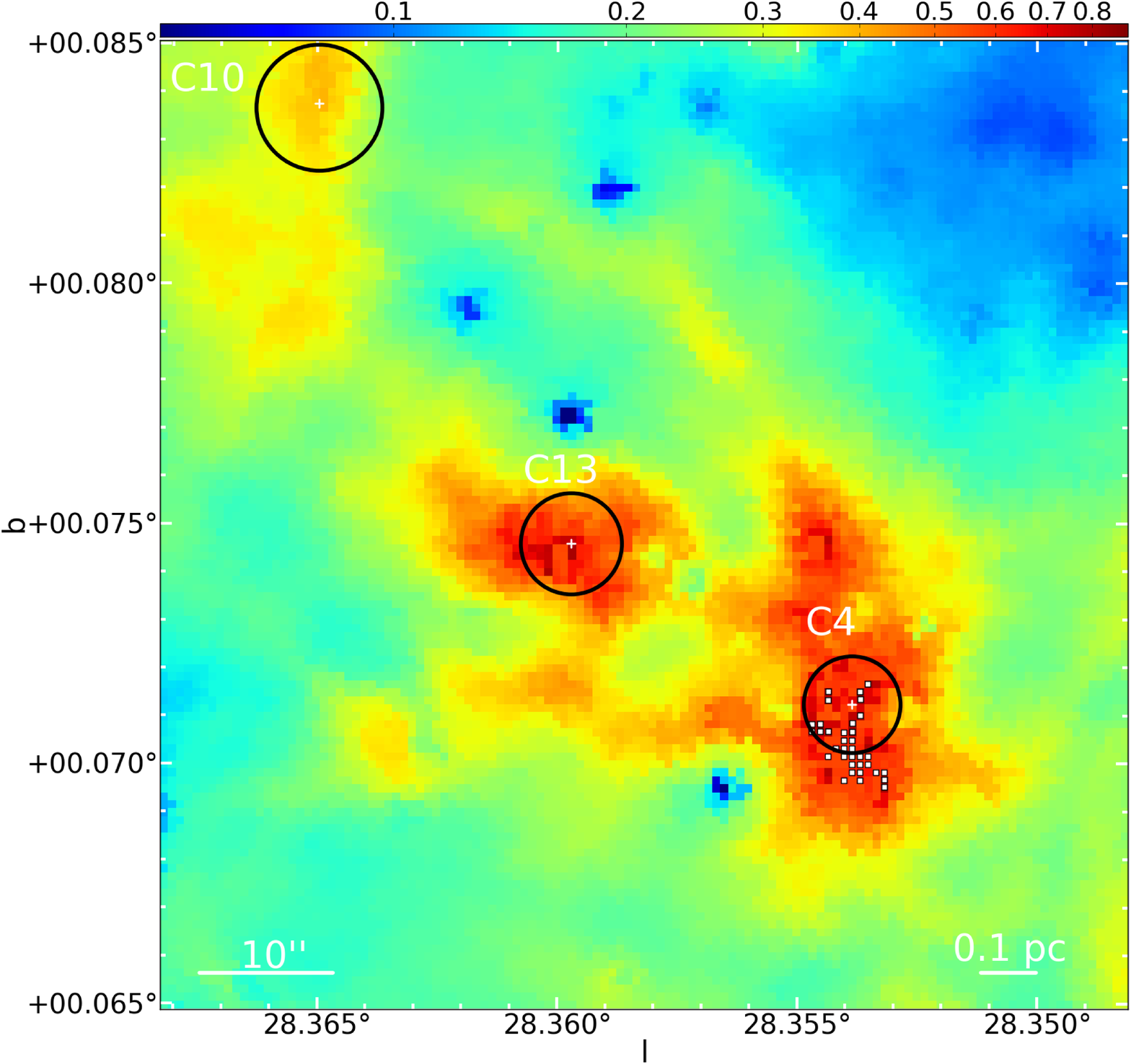} & \hspace{-0.0in} \includegraphics[width=1.9in]{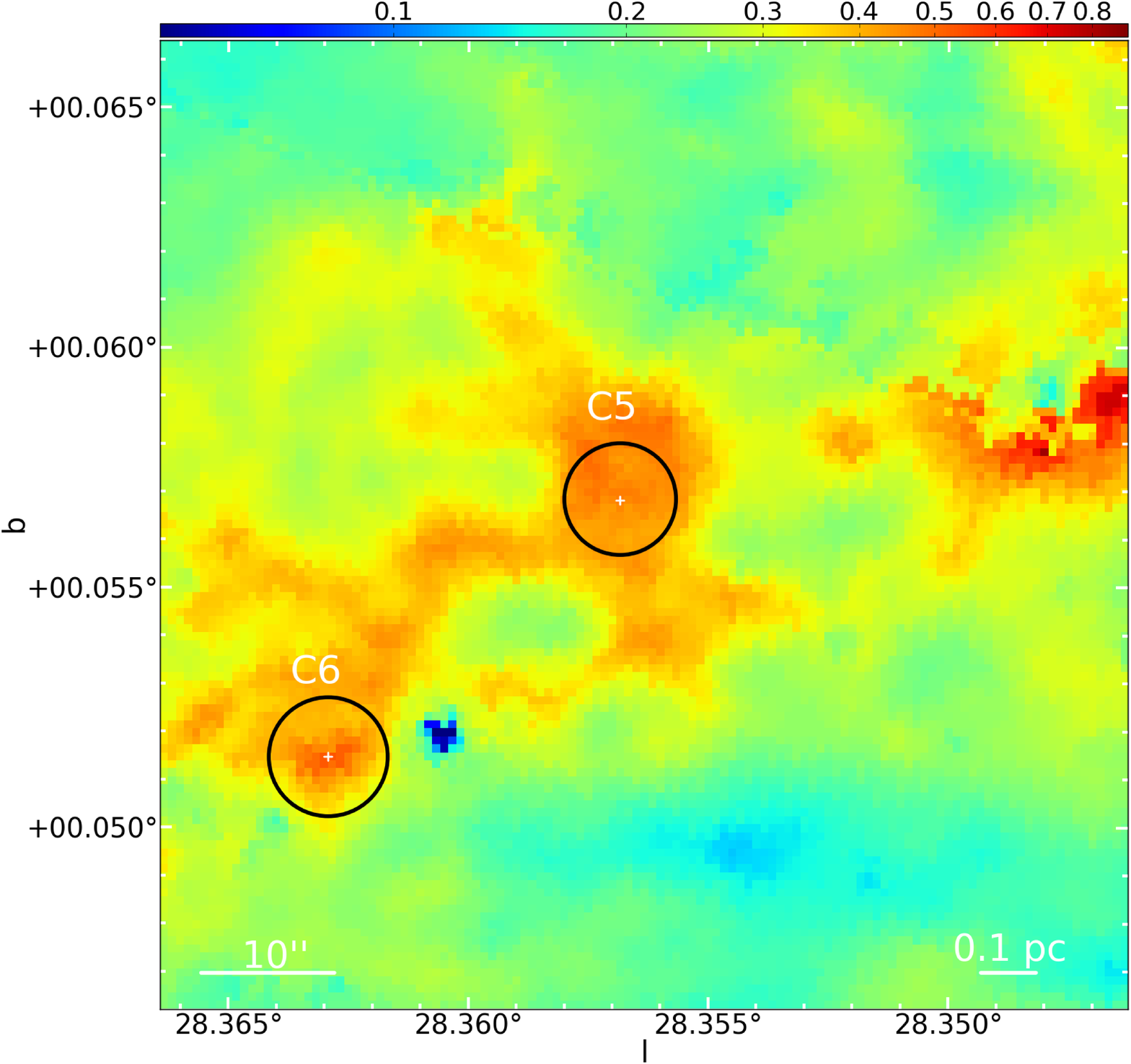} & \hspace{-0.0in} \includegraphics[width=1.9in]{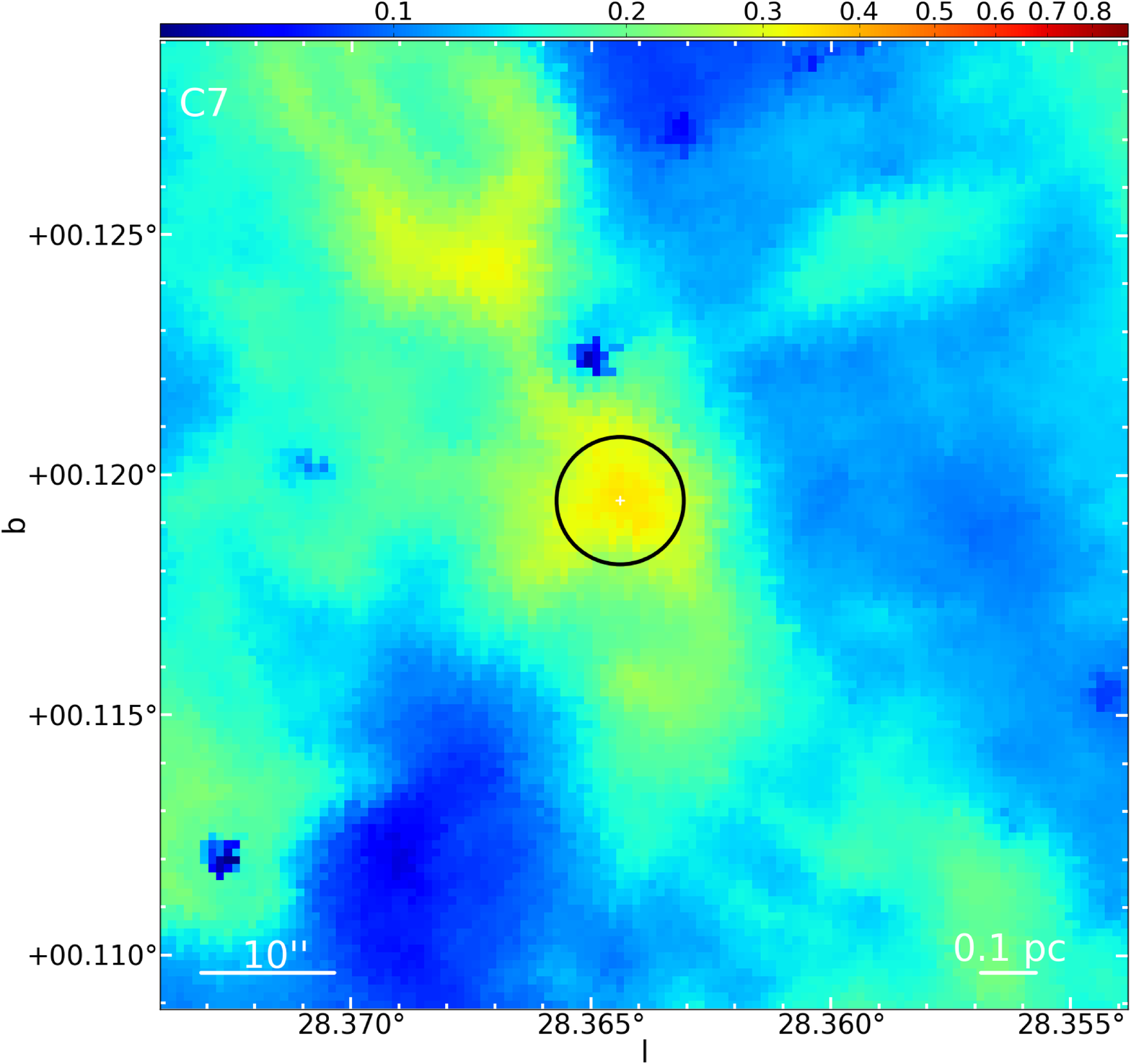} \\
\hspace{-0.0in} \includegraphics[width=1.9in]{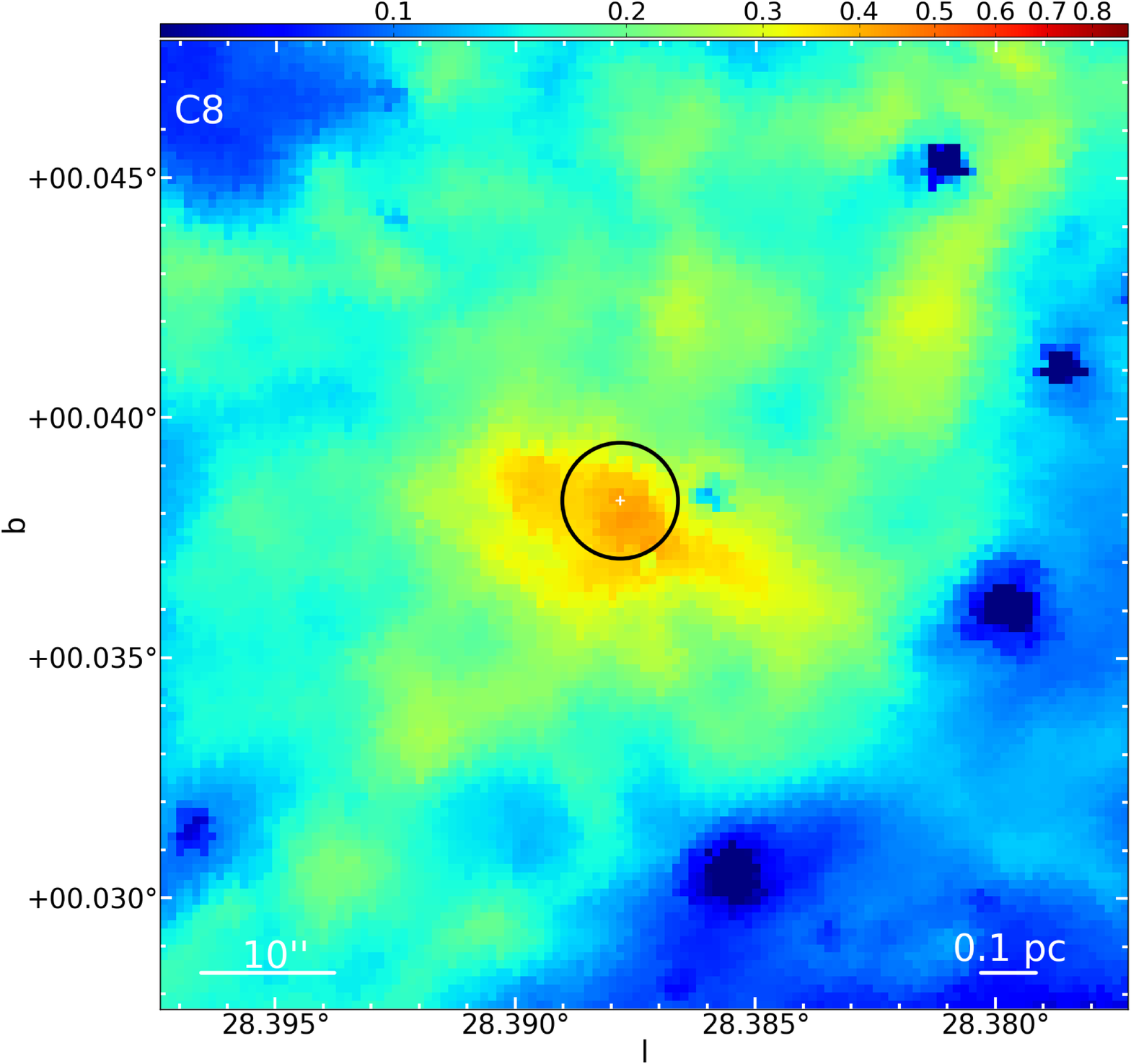} & \hspace{-0.0in} \includegraphics[width=1.9in]{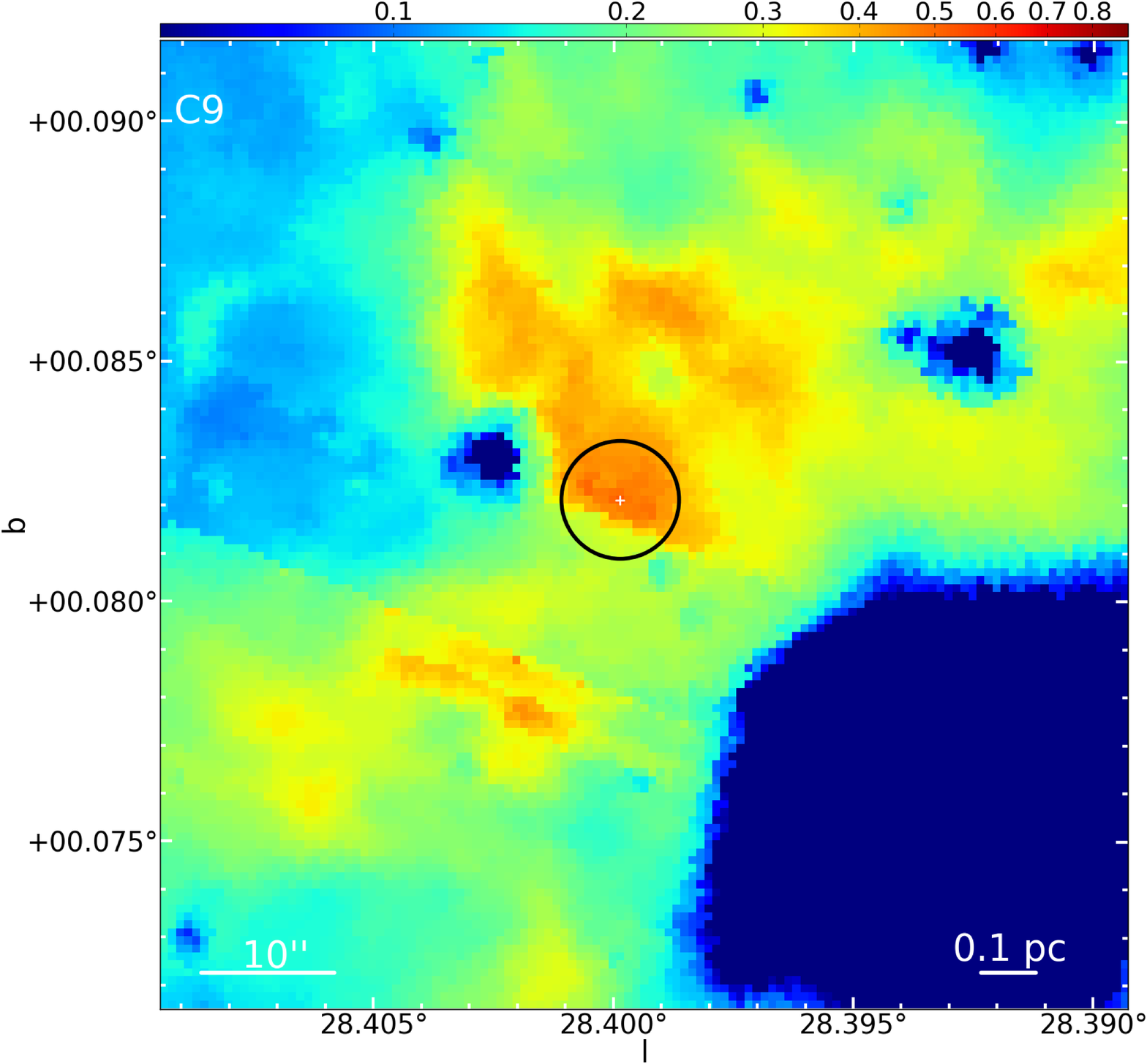} & \hspace{-0.0in} \includegraphics[width=1.9in]{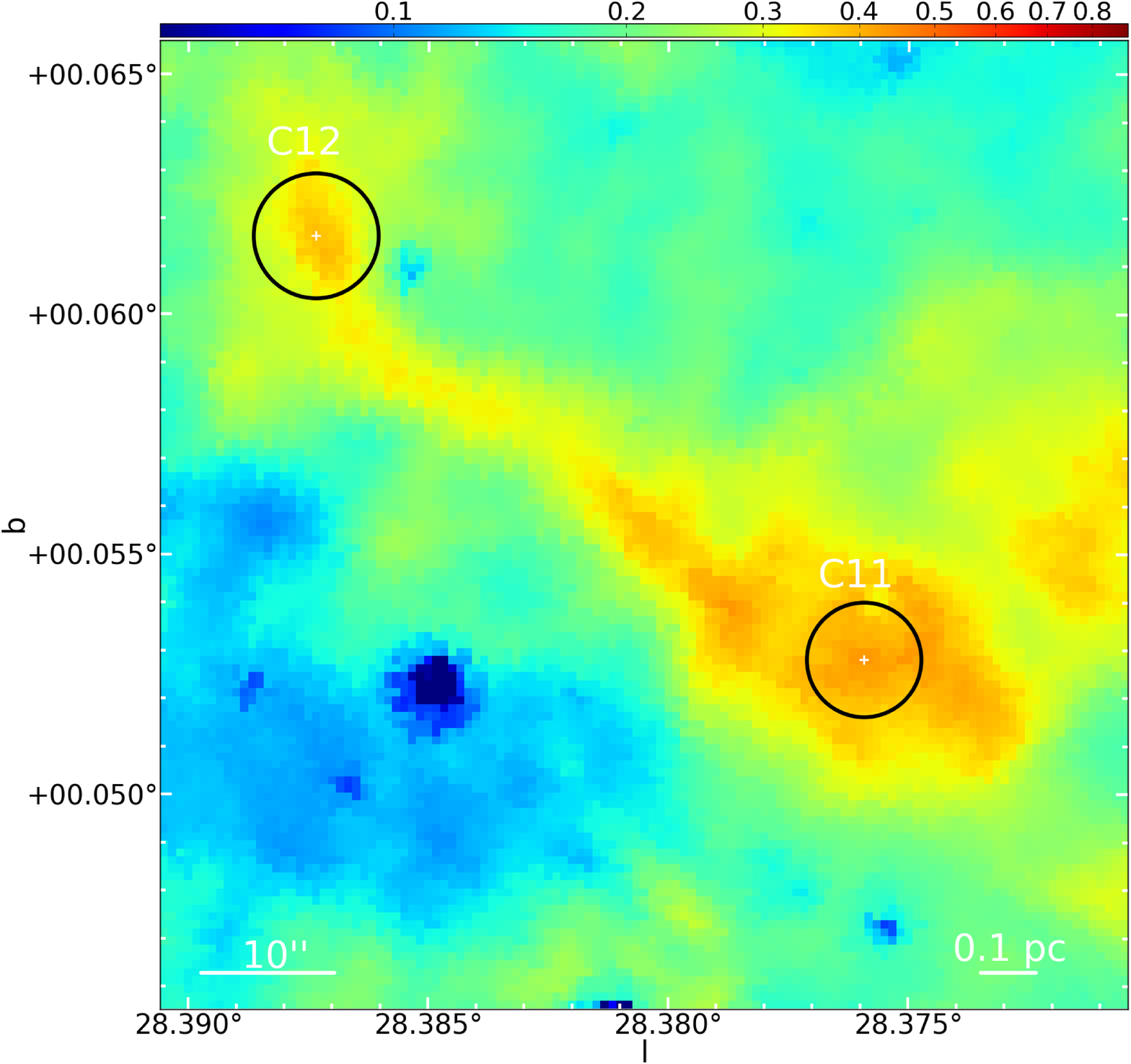} \\
\end{array}$
\end{center}
\caption{\footnotesize
Zoom images of 9 regions from Fig.~1b, illustrating 16 identified
core/clumps. {\it Top row, left to right:} (a) C1, C16; (b) C2, C14,
C15; (c) C3. {\it Middle row, left to right:} (d) C4, C10, C13;
(e) C5, C6; (f) C7. {\it Bottom row, left to right:} (g) C8; (h)
C9; C11, C12. For each core/clump, the black circle shows the region
enclosing $60\:M_\odot$. White squares mark saturated pixels, present
in C2 and C4 (see text).
\label{fig:cores}
}
\end{figure*}


The properties of 9 dense core/clumps (C1-C9) within the IRDC were
studied by \citet{butler2012}.
We now expand the sample, to obtain a more complete census of the
massive starless or early-stage core/clumps in the cloud, identifing 7
new core/clumps (C10-C16): these are the 7 highest $\Sigma$ local
maxima external to C1-C9. They are also required to be free of
$8\:\micron$ sources.
Zoomed-in views of the cores are presented in Figure~\ref{fig:cores}.

Only C2 and C4 exhibit saturation as defined in
\S\ref{S:method}. Following \citet{butler2012}, we use the unsaturated portions of
the $\Sigma$ profiles to derive the power law volume density structure
of the core/clumps, $\rho_{\rm{cl}}(r)=\rho_{\rm{s,cl}}(r/R_{\rm{cl}})^{-k_{\rho,{\rm{cl}}}}$,
where $\rho_{\rm{s,cl}}=\mu_{\rm{H}}n_{\rm{H,s,cl}}$ (with
$\mu_{\rm{H}}=2.34\times10^{-24}\:{\rm{g}}$) is the density at the
surface of the core/clump, $R_{\rm{cl}}$. We hereafter refer to these
objects as ``clumps''. We also consider background-subtracted
$\Sigma$ profiles, where a uniform average background is evaluated
from the annulus from $R_{\rm{cl}}$ to $2R_{\rm{cl}}$, and describe
these objects as ``cores''.

We first consider the properties of the core/clumps at a scale that
encloses $M_{\rm{cl}}=60\:M_\odot$, which is approximately the mass
scale needed to form a massive (O) star \citep{mckee2003}. We find the
mean/median/dispersion of the $k_{\rho,{\rm{cl}}}$ values are
0.826/0.775/0.262. Equivalently, for the envelope-subtracted cores we
find $k_{\rho,c}$ mean/median/dispersion values of
1.33/1.39/0.299. These are similar to the \citet{butler2012} results.  The mean core
mass is $M_c=14.9\:M_\odot$, so the method of envelope subtraction
typically removes about 3/4 of the total mass that is projected along
the line of sight. Envelope subtraction depends on the assumed 3D
geometry of the core and its surrounding clump and so is quite
uncertain. Astrochemical tracers of starless cores, such as
$\rm{N_2D^+}$, are thus useful for helping to locate these objects
more precisely \citep{tan2013}.

The mean/median/dispersion values for $\Sigma$ of the $60\:M_\odot$
clumps are 0.446/0.424/0.109~$\gcc$ and for the cores are
0.118/0.107/0.0938~$\gcc$. If core properties are better estimated
after envelope subtraction, then their $\Sigma$'s are relatively
small, e.g. compared to fiducial values of the turbulent core model of
massive star formation \citep{mckee2003} with $\Sigma\sim1\:\gcc$ or the
hypothesized minimum $\Sigma$ threshold for massive star formation of
$1\:\gcc$ due to accretion powered heating suppression of
fragmentation \citet{krumholz2008}. Even if the maximum values,
i.e. with no envelope subtraction, are assigned to the cores, then
these are still somewhat smaller than $1\:\gcc$. Note, that these 16
cores are amongst the highest $\Sigma$ peaks in the extinction
map. Note also, that at least some (e.g., C1) and probably most of the
cores are very cold ($\sim 10-15$~K) (Wang et al. 2008; Tan et
al. 2013), so radiative heating is not currently suppressing their
fragmentation. Magnetic suppression of fragmentation of these
core/clumps, requiring $\sim0.1-1\:$mG fields, remains a possibility
\citep{butler2012,tan2013}.

The structural properties of the core/clumps at the
$M_{\rm{cl}}=60\:M_\odot$ scale are summarized in
Table~\ref{tab:coresample}. Also shown here are the results of a more
general fitting of power-law density profiles where the outer radius
is varied and the reduced $\chi^2$ of the fit minimized
\citep{butler2012}. Sometimes the best-fit model is quite close the
$M_{\rm{cl}}=60\:M_\odot$ model, but in other cases it can shrink
somewhat or grow up to $\sim10^3\:M_\odot$. The roughly similar
structural properties over a range of length and mass scales may
indicate that there is a self-similar hierarchy of structure present
in the IRDC. At the same time we caution that the ``best-fit'' model
is often not that much better than models fit to other
scales. Nonspherical geometries can also be important, which can limit
the applicability of these simple spherical models.

\begin{deluxetable}{cccccccccccc}
\tabletypesize{\tiny}
\tablecolumns{12}
\tablewidth{0pt}
\tablecaption{Structural Properties of Core/Clumps\tablenotemark{a}}
\tablehead{
\colhead{Name} & \colhead{l} & \colhead{b} & \colhead{$R_{\rm{cl}}=R_c$} & \colhead{$\bar{\Sigma}_{\rm cl}$} & \colhead{$\bar{\Sigma}_{c}$} & \colhead{$k_{\rm \rho,cl}$} & \colhead{$k_{\rho,c}$} & \colhead{$n_{\rm H,s,cl}$} & \colhead{$n_{\rm H,s,c}$} & \colhead{$M_{\rm cl}$} & \colhead{$M_{\rm c}$\tablenotemark{b}}\\
\colhead{} & \colhead{($^\circ$)} & \colhead{($^\circ$)} & \colhead{(pc)} & \colhead{($\rm g\:cm^{-2}$)} & \colhead{($\rm g\:cm^{-2}$)} & \colhead{} & \colhead{} & \colhead{($\rm 10^5 cm^{-3}$)} & \colhead{($\rm 10^5 cm^{-3}$)} & \colhead{($M_\odot$)} & \colhead{($M_\odot$)}
}
\startdata
C1 & 28.32450 & 0.06655 & 0.0806 & 0.613 & 0.141 & 1.06 & 1.53 & 5.84 & 0.951 & 60.0 & 13.7\\
   &          &         & 0.0581 & 0.637 & 0.151 & 1.06 & 1.58 & 5.38 & 1.27  & 32.4 & 7.70\\
\hline                                           
C2 & 28.34383 & 0.06017 & 0.0793 & 0.633 & 0.254 & 1.18 & 1.86 & 6.18 & 2.13  & 60.0 & 24.1\\                      
   &          &         &   0.0872      &     0.630      &       0.246       &     1.14      &    1.50    &   3.75    &    1.46     &   72.2    &   28.2\\
\hline
C3 & 28.35217 & 0.09450 &   0.120       &     0.276      &       0.0200      &     0.775     &    1.18    &    1.51   &    0.0822   &   60.0    &   4.36\\
   &          &         &   0.0581      &     0.287      &       0.0257      &     0.740     &    1.36    &   2.80    &   0.251     &   14.6    &   1.31\\                    
\hline
C4 & 28.35417 & 0.07067 &   0.0806      &     0.614      &       0.335       &     1.36      &    1.95    &    6.04   &    2.67     &   60.0    &   32.7\\
   &          &         &   0.101       &     0.607      &       0.274       &     1.36      &    1.82    &   2.43    &    1.10     &   94.6    &   42.8\\
\hline
C5 & 28.35617 & 0.05650 &   0.0942      &     0.449      &       0.0575      &     0.538     &    0.979   &   3.73    &    0.371    &   60.0    &   7.67\\
   &          &         &   0.0872      &     0.453      &       0.0670      &    0.5200     &    0.600   &   4.31    &    0.637    &   51.8    &   7.70\\
\hline
C6 & 28.36267 & 0.05150 &   0.100       &     0.392      &       0.179       &    0.777      &    1.42    &   3.28    &     1.28    &   60.0    &   27.4\\
   &          &         &   0.843       &      0.315     &        0.0951     &     0.800     &     1.14   &   0.240   &    0.0725   &    3370   &    1020\\
\hline
C7 & 28.36433 & 0.11950 &   0.108       &     0.338      &      0.0494       &    0.759      &    1.08    &   2.26    &    0.247    &  60.0     &   8.75\\
   &          &         &   0.0581      &     0.356      &      0.0558       &    0.680      &    1.14    &   3.94    &    0.616    &  18.1     &   2.84\\
\hline
C8 & 28.38783 & 0.03817 &   0.0979      &     0.415      &       0.226       &     1.10      &    1.57    &   2.87    &    1.21     &  60.0     &   32.6\\
   &          &         &   0.218       &     0.368      &       0.143       &    1.02       &    1.70    &  0.759    &   0.296     &  263      &   103\\
\hline
C9 & 28.39950 & 0.08217 &   0.0997      &     0.401      &      0.0496       &    0.707      &    1.14    &  3.64     &   0.331     &  60.0     &   7.43\\
   &          &         &   0.0872      &     0.405      &      0.0553       &    0.740      &    1.14    &   2.98    &   0.408     &  46.3     &   6.34\\
\hline
C10 & 28.36486 & 0.08397 &  0.107       &     0.346      &      0.0456       &    0.737      &    1.23    &   2.66    &   0.235     &  60.0     &   7.90\\
    &          &         &  0.0872      &     0.351      &      0.0508       &    0.480      &    1.12    &   2.61    &   0.378     &  40.2     &   5.82\\
\hline
C11 & 28.37600 & 0.05279 &  0.0968      &     0.424      &      0.107        &    0.563      &    0.920   &   3.47    &   0.729     &  60.0     &   15.1\\
    &          &         &  0.130       &     0.421      &      0.100        &    0.520      &    0.980   &   2.24    &   0.536     &  108.     &   25.9\\
\hline
C12 & 28.38717 & 0.06128 &  0.106       &     0.352      &      0.0450       &    0.900      &    1.42    &   2.62    &   0.279     &  60.0     &   7.66\\
    &          &         &  0.0581      &     0.370      &      0.0564       &    0.860      &    1.64    &   2.99    &   0.455     &  18.8     &   2.87\\
\hline
C13 & 28.33317 & 0.05900 &  0.0846      &     0.557      &      0.180        &    1.04       &    1.47    &   5.2     &   1.40      &  60.0     &   19.4\\
    &          &         &  0.0872      &     0.555      &      0.198        &    1.00       &    1.56    &   3.17    &   1.13      &  63.6     &   22.7\\
\hline
C14 & 28.33401 & 0.06383 &  0.103       &     0.374      &      0.0408       &    0.383      &    1.07    &   2.87    &   0.261     &  60.0     &   6.54\\
    &          &         &  0.130       &     0.372      &      0.0445       &    0.360      &    0.720   &   2.24    &   0.268     &  95.9     &   11.4\\
\hline
C15 & 28.33418 & 0.06366 &  0.0956      &     0.435      &      0.136        &    0.621      &    1.39    &   3.60    &    0.938    &  60.0     &   18.7\\
    &          &         &  0.290       &     0.409      &      0.0994       &    0.160      &    1.00    &   0.974   &    0.236    &   521     &   126\\
\hline
C16 & 28.32985 & 0.06717 &  0.0878      &     0.516      &      0.0350       &    0.710      &    1.09    &   4.65    &    0.263    &   60.0    &   4.07\\
    &          &         &  0.0581      &     0.521      &      0.0344       &    0.620      &    1.10    &   5.88    &    0.389    &   26.5    &   1.75\\
\hline
\hline                                                                                                        
Mean &         &         &  0.0963      &     0.446      &      0.118        &    0.826      &    1.33    &   3.78    &   0.836     &  60.0     &   14.9\\
    &          &         &  0.152       &     0.441      &      0.106        &    0.754      &    1.26    &   2.92    &   0.594     &  302      &   88.5\\
\hline
Median &       &         &  0.0979      &     0.424      &      0.107        &    0.775      &    1.39    &   3.60    &   0.729     &  60.0     &   13.7\\
       &       &         &  0.0872      &     0.409      &      0.0951       &    0.740      &    1.14    &   2.98    &   0.455     &  63.6     &   11.4\\
\hline
Dispersion &   &         &  0.0115      &     0.109      &      0.0938       &    0.262      &    0.299   &   1.41    &   0.750     &  0.00     &   9.84\\
        &      &         &  0.195       &     0.114      &      0.0763       &    0.313      &    0.353   &   1.54    &   0.417     &  828      &   251\\
\enddata
\tablenotetext{a}{First line for each source shows results at the scale where the total projected enclosed mass is $M_{\rm{cl}}=60\:M_\odot$; second line shows results at the ``best-fit'' scale, minimizing the reduced $\chi^2$ of the projected power law density fit to the $\Sigma(r)$ profile.}
\tablenotetext{b}{Based on fitted power law density profile, which compensates for mass missed in saturated core centers, i.e. relevant for C2 and C4, typically $\lesssim 10\%$ of total.}
\label{tab:coresample}
\end{deluxetable}

\subsection{The Mass Surface Density Probability Distribution Function}

The $\Sigma$ (or $A_V$ or $N_{\rm{H}}$) PDF, either area
($p_A(\Sigma)$) or mass-weighted ($p_M(\Sigma)$), entrains information
about cloud self-gravity, turbulence, shocks and magnetic field
support. Our temperature-independent high-dynamic range extinction map
can yield the best constraints on the $\Sigma$ PDF. However, the
relatively limited areal coverage of the archival {\it Spitzer} data
require us to utilize the GLIMPSE-based maps of \citet{kainulainen2013} (offset-corrected
to smoothly join the new map) to extend the region to completely cover
a 20\arcmin$\times$19\arcmin rectangular area that encloses an
$A_V=3\:$mag contour \citep[cf. the 15\arcmin$\times$15\arcmin\ area
enclosing the $A_V=7\:$mag contour of][]{kainulainen2013}.

\begin{figure*}[!tb]
\begin{center}$
\includegraphics[width=2.8in]{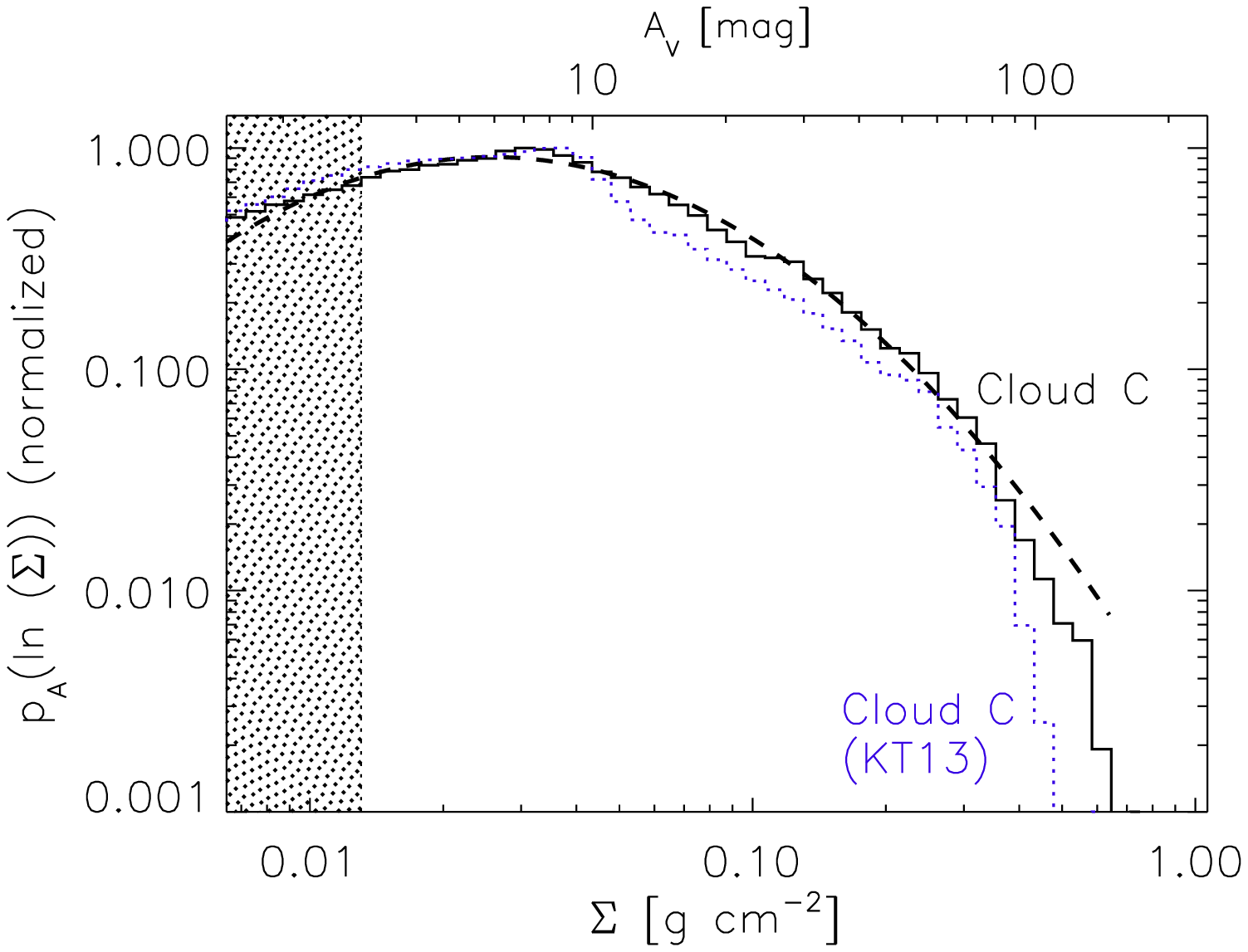}
\includegraphics[width=2.8in]{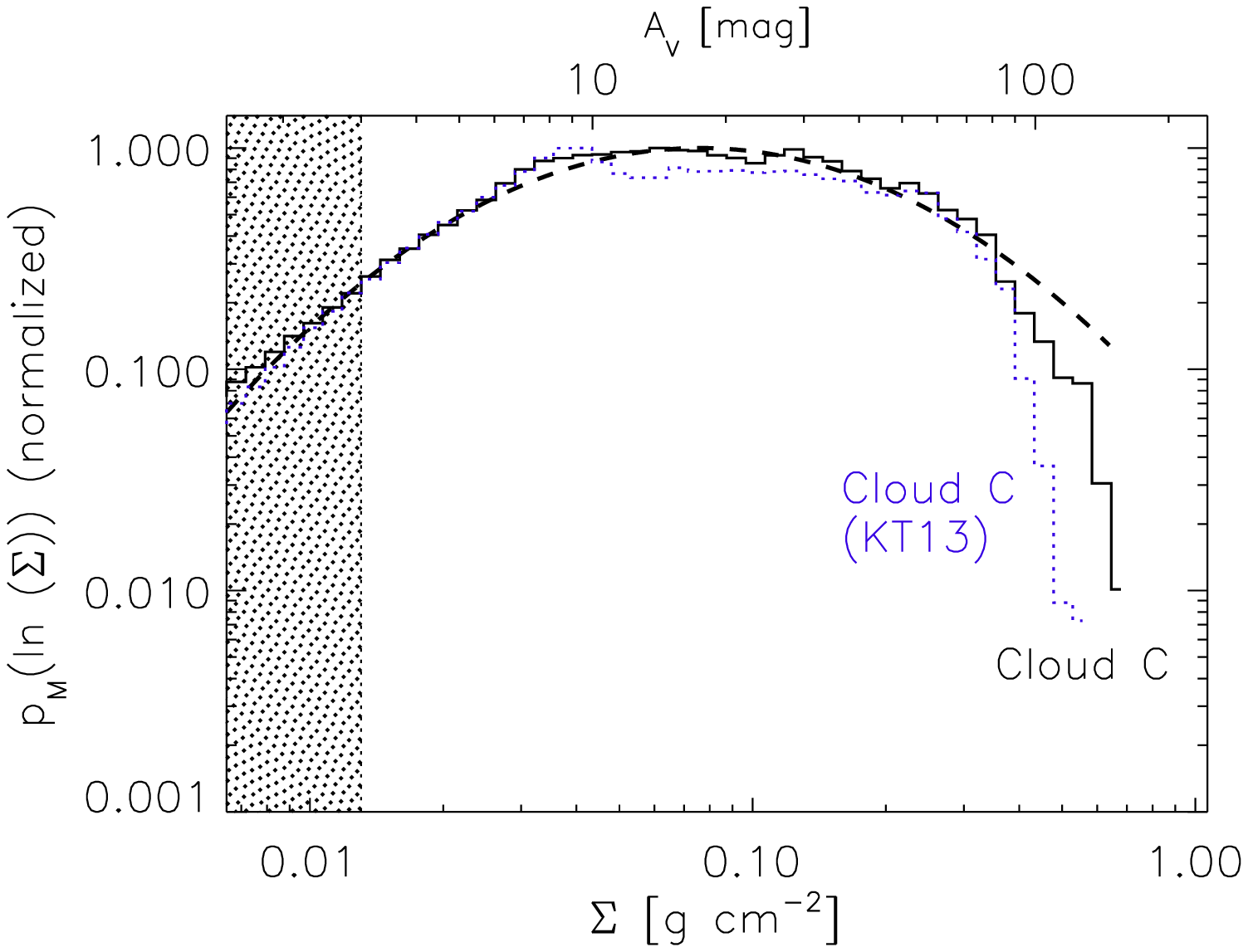}
$
\end{center}
\caption{
{\it{(a) Left:}} Area-weighted probability distribution function of
$\Sigma$ for IRDC G028.37+00.07 (Cloud C) (black histogram) and
best-fit log-normal (dashed line) covering a contiguous region
completely enclosing an $A_V>3\:$mag contour --- above this level,
shown by the shaded/nonshaded boundary, the PDF is complete, modulo
saturation effects for $A_V\gtrsim200\:$mag. We also show the
\citet{kainulainen2013}-derived PDF (purple histogram), based on a shallower, smaller
dynamic range extinction map. {\it{(b) Right:}}
Mass-weighted PDFs, with same notation as (a).
}
\label{fig:pdf}
\end{figure*}

The area and mass-weighted $\Sigma$ PDFs are shown in
Fig.~\ref{fig:pdf}. 
The new PDF extends to both lower and higher values of $\Sigma$
compared to that derived by \citet{kainulainen2013}. 
We fit a log-normal
function to $p_A({\rm{ln}}\Sigma)$, and then use this to derive
$\overline{\rm{ln}\Sigma}$ (over the considered range of $\Sigma$),
which then defines the mean
$\overline{\Sigma}_{\rm{PDF}}\equiv{e}^{\overline{\rm{ln}\Sigma}+\sigma^2_{\rm{ln}\Sigma}/2}$,
where $\sigma_{\rm{ln}\Sigma}$ is the standard deviation of
$\rm{ln}\Sigma$ \citep[note typo in eq.~26 of][]{kainulainen2013}, used to define the
  mean-normalized mass surface density,
  $\Sigma^\prime\equiv\Sigma/\overline{\Sigma}_{\rm{PDF}}$. For this
  distribution, we find the best-fit log-normal
\begin{equation}
p_A({\rm ln}\Sigma^\prime) =\frac{1}{(2\pi)^{1/2}\sigma_{\rm{ln}\Sigma^\prime}}
{\rm{exp}}\left[-\frac{({\rm{ln}}\Sigma^{\prime}-\overline{{\rm{ln}}\Sigma})^2}{2\sigma_{\rm{ln}\Sigma^\prime}^2}\right],
\end{equation}
where  
$\sigma_{\rm{ln}\Sigma^\prime}$ is the standard deviation of
$\rm{ln}\Sigma^\prime$. We find
$\overline{\Sigma}_{\rm{PDF}}=0.039\:\gcc$
(i.e. $\overline{A}_{V,{\rm{PDF}}}=9.0\:$mag) and
$\sigma_{\rm{ln}\Sigma^\prime}=1.4$, compared with $8.3\:$mag and 1.7
found by \citet{kainulainen2013}, respectively. Note, \citet{kainulainen2013} only fit to $A_V>7\:$mag,
insufficient to determine the PDF peak so the parameters of their
log-normal fit are less accurate.

The derived PDF is well fit by a single log-normal.  Deviation at high
$\Sigma\gtrsim0.5\:\gcc$ may be due to saturation (\S\ref{S:method})
or absence of MIR-bright high $\Sigma$ regions. The value of
$\overline{A}_{V,{\rm{PDF}}}$ is much higher than in nearby
star-forming clouds ($\overline{A}_{V,{\rm{PDF}}}\simeq0.6-3.0\:$mag)
\citep{kainulainen2009}. G028.37+00.07 also has
a higher dense gas fraction than other studied IRDCs \citep[e.g.,][]{kainulainen2013}.
The PDF shows no indication of a high-end power law tail (indeed there is
no room for such a tail if the PDF peak is dominated by the
log-normal), observed in some clouds \citep[e.g.,][]{kainulainen2011} and modeled as being
due to a separate self-gravitating component
\citep[e.g.,][]{kritsuk2011} \citep[however, see][]{kainulainen2011b}.
Since the observed line-widths (\S\ref{S:dynamics}) indicate the
overall cloud is self-gravitating, this may imply that a self-similar,
self-gravitating hierarchy of structure is present over the complete
range of spatial scales in the cloud probed by our extinction
map and that such a hierarchy produces a log-normal-like $\Sigma$ PDF
\citep[see also][]{goodman2009}. This needs to be explored in global
(non-periodic box) simulations of magnetized, self-gravitating
molecular clouds.





\acknowledgements We thank the referee for helpful comments. We acknowledge
NASA grant ADAP10-0110 (JCT) and Deutsche Forschungsgemeinschaft program 1573 (JK).

\end{document}